\documentclass{aa}  
\usepackage[utf8]{inputenc}
\usepackage[T1]{fontenc}
\usepackage{threeparttable}
\usepackage{amssymb}
\usepackage{amsmath}
\usepackage{kantlipsum}
\usepackage{multicol}
\usepackage[printwatermark]{xwatermark}
\usepackage{array}
\usepackage{color}
\usepackage{url}
\usepackage{enumitem}
\usepackage{bm}
\usepackage{xcolor}
\usepackage{afterpage}
\usepackage[normalem]{ulem}
\usepackage{graphicx}
\usepackage{stfloats}
\usepackage{txfonts}
\usepackage{caption}
\usepackage{balance}
\usepackage{ragged2e}
\usepackage[rightcaption]{sidecap}

\usepackage{savesym}
\savesymbol{iint}
\usepackage{txfonts}
\restoresymbol{TXF}{iint}

\begin{document} 

\newcommand{\Tp}{T_{p}}
\newcommand{\Rp}{R_{p}}
\newcommand{\Rs}{R_{\ast}}
\newcommand{\Mb}{M_{b}}
\newcommand{\Ms}{M_{*}}
\newcommand{\Teff}{T_\mathrm{eff}}
\newcommand{\Msun}{\textrm{M}_\mathrm{\odot}}
\newcommand{\Mearth}{\textrm{M}_\mathrm{\oplus}}
\newcommand{\Mj}{\textrm{M}_\mathrm{J}}
\newcommand{\Rj}{\textrm{R}_\mathrm{J}}

\newcommand{\ee}{\end{equation}}

   \title{Circumbinary exoplanets and brown dwarfs with the Laser
Interferometer Space Antenna}


   \author{C. Danielski
          \inst{1,2}
          \and
          V. Korol \inst{3,4}
          \and
          N. Tamanini \inst{5}
          \and
          E. M. Rossi \inst{3}
          }

   \institute{AIM, CEA, CNRS, Universit\'e Paris-Saclay, Universit\'e Paris Diderot, Sorbonne Paris Cit\'e, F-91191 Gif-sur-Yvette, France. 
    \email{camilla.danielski@cea.fr}
    \and Institut d'Astrophysique de Paris, CNRS, UMR 7095, Sorbonne Universit\'e, 98 bis bd Arago, 75014 Paris, France
    \and Leiden Observatory, Leiden University, PO Box 9513, 2300 RA, Leiden, the Netherlands
    \and School of Physics and Astronomy, University of Birmingham, Edgbaston, Birmingham B15 2TT, United Kingdom
    \and Max-Planck-Institut f\"ur Gravitationsphysik, Albert-Einstein-Institut, Am M\"uhlenberg 1,14476 Potsdam-Golm, Germany.}


 
  \abstract
   {}
   {
   We explore the prospects for the detection of giant circumbinary exoplanets and brown dwarfs (BDs) orbiting Galactic double white dwarfs binaries (DWDs) with the Laser Interferometer Space Antenna (LISA).}
   %
   {By assuming an occurrence rate of 50$\%$, motivated by white dwarf pollution
   observations, we built a Galactic synthetic population of P-type giant exoplanets and BDs orbiting DWDs. 
   We carried this out by injecting different sub-stellar populations, with various mass and orbital separation characteristics, 
   into the DWD population used in the LISA mission proposal. We then performed a Fisher matrix analysis to measure how many of these three-body systems show a periodic Doppler-shifted gravitational wave perturbation detectable by LISA.}
   {We report the number of circumbinary planets (CBPs) and (BDs) that can be detected by LISA for various 
   combinations of mass and semi-major axis distributions. We identify pessimistic and optimistic scenarios corresponding, respectively, to 3 and 83 (14 and 2218) detections of CBPs (BDs), observed during the length of the nominal LISA mission.
   These detections are distributed all over the Galaxy following the underlying DWD distribution, and they are biased towards DWDs with higher LISA signal-to-noise ratio and shorter orbital period.
   Finally, we show that if LISA were to be extended for four more years, the number of systems detected will be more than doubled in both the optimistic and pessimistic scenarios. 
   }
   {Our results present promising prospects for the detection of post-main sequence exoplanets and BDs, 
   showing that gravitational waves can prove the existence of these populations over the totality of the Milky Way.
   Detections by LISA will deepen our knowledge on the life of exoplanets subsequent to the most extreme evolution phases of 
   their hosts, clarifying whether new phases of planetary formation take place later in the life of the stars.
   Such a method is strongly complementary to electromagnetic studies within the solar region and opens
   a window into the investigation of planets and BDs everywhere in the entire Galaxy, and possibly even in nearby galaxies in the Local Group.\\
   }


   \maketitle
%

\section{Introduction}
\label{sec:intro}
In an epoch in which 
the field of exoplanets is moving at a fast pace and groundbreaking discoveries are made, very little is known about the ultimate fate of planetary systems.
In the Milky Way more than $\sim$ 97\% of the stars will turn into a white dwarf (WD), meaning that the vast majority of the more than known 3000 planet-hosting stars will end their life as WDs. Can their planets survive stellar evolution? 
Theoretical models indicate that a planet can endure the host-star evolution if it avoids engulfment or evaporation throughout the red giant or/and the asymptotic giant branch phases  (e.g. \citealt{LivioSoker1984, DuncanLissauer1998,nel98}), where survival itself depends, among various parameters, on the initial semi-major axis and planetary mass \citep{VillaverLivio2007}.
For what remains of the planetary system the complex long-term orbital evolution, consequent to stellar evolution, may yield to planet ejections and/or collisions (e.g.  \citealt{DebesSigurdsson2002, Veras2011, Veras2016, Mustill2018}).
Besides, if migration or scattering occurs towards the proximity of the Roche limit,
strong tidal forces can further crush the planetary cores \citep{Farihi2018}, like in the case of the planetesimal found shattering around  WD 1145+017 \citep{Vanderburg2015}.
Such a fragmentation process consequently enables the formation of a debris disc, made of metal-rich planetary material, which could in turn 
accrete onto the WD, $polluting$ its atmosphere (e.g.  \citealt{jura2009, Farihi2010, Farihi2016, Veras2016, Brown2017, Smallwood2018}).
\section{Evidences for sub-stellar objects around WDs}
\subsection{White dwarf pollution}
\label{sec:wdpollution}
White dwarfs are expected to have a pure H or He atmosphere \citep{Schatzman1945} and their high surface gravity ($\sim 10^5$ denser than the Sun) makes the sinking metals diffusion timescale several order of magnitude shorter than the evolutionary period.
Yet, observations show the presence of heavy elements in the spectra of 25$\%$ to 50$\%$ of all observed WDs  \citep{Zuckerman2003, Zuckerman2010, Koester2014}, indicating that a continuous supply of metal-rich material accreting onto these  stars must be present. 
There are several sources of WD pollution proposed in the literature.
\ This WD pollution could originate from planetary material (i.e. from circumstellar debris discs as previously explained), moons via planet-planet scattering \citep{Payne2016, Payne2017}, or comets \citep{CaiazzoHeyl2017}. 
It could also be from perturbations created by eccentric high-mass planets, which drive substantial
asteroids or minor bodies to the innermost orbital region around the star  
(which in some cases is within the stellar Roche limit), 
thereby yielding to tidal fragmentation (e.g.  \citealt{Frewen2014, Chen2019, Wilson2019}). 
The last source of WD pollution is currently preferred in the community. 
Pollution of WDs in wide binaries may also be caused by Kozai-Lidov instabilities, which can cause the orbit of objects such as planets, to intersect the tidal radius of the WD, causing their distruction \citep{HamersPortegies2016, PetrovichMunoz2017}.

Overall WD pollution studies support the evidence of dynamically active planetary systems orbiting WDs. Nonetheless, because of the intrinsic low luminosity of these stars, no planets have been detected yet around single WDs, but an intact planetesimal has been observed inside a debris disc belonging to a WD \citep{Manser2019}.

\subsection{Generations of circumbinary post-common envelope exoplanets}
\label{sec:postCEexoplanets}

Contrary to the single star case, P-type exoplanets \citep{Dvorak1986} have been detected orbiting binary stars in which the higher mass component has already grown to be a WD (i.e. the mass of its progenitor is $M_* \lesssim$ 10 $\Msun$); the second component is usually a low-mass star that will become a giant later in its life
(i.e. NN Ser, HU Aqr, RR Cae, UZ For, and DP Leo; \citealt{Beuermann2010A, Qian2011, Qian2012, Potter2011, Qian2010, Beuermann2011}).\
These discoveries prove that planets can survive at least one common envelope (CE) phase, i.e.~a shared stellar atmosphere phase typical of close binary stars, which happens when one of the binary components becomes a giant (see Section \ref{sec:dwds} for more detail on this phase). Surviving planets (in this case first generation planets) are usually called  post-main sequence exoplanets or  post-CE exoplanets, and they are more likely to survive around evolving close binary stars than around evolving single stars \citep{Kostov2016}. Only a small amount is known about these planets,  
but they are extremely interesting as they provide a link between planetary formation and fate, as well as constraints on tidal, binary mass loss, and radiative process \citep{Veras2016}. 

Detection and study of these bodies can also provide us with further information about planetary formation processes.
There is the interesting hypothesis that some of these known post-CE planets
belong to a `new generation', i.e. they have formed after the first CE phase 
(e.g.  \citealt{ZorotovicSchreiber2013, Volschow2014}). 
A study by \cite{KashiSoker2011} has shown that because of angular momentum conservation and further interaction with the binary system, 1 to 10$\%$ of the ejected envelope does not reach the escape velocity. This material remains bound to the binary system, falls back on it, flattens, and forms a circumbinary disc, which could provide the necessary environment for the formation of a second generation of massive exoplanets \citep{Perets2010, Volschow2014, SchleicherDreizler2014}.
On the other hand, as already mentioned for the single star case, some sub-stellar bodies (e.g.  first generation exoplanets, asteroids, and comets), whose orbit is small and/or eccentric enough, could be tidally disrupted during the CE phase, creating a circumbinary disc of rocky debris, out of which new terrestrial exoplanets can grow \citep{Farihi2017}.
In both cases photo-heating from the binary, photoionisation, radiation pressure, and differences in the magnetic field, would likely be responsible for influencing the discs in different ways, causing second generation planets to differ from first generation planets \citep{Perets2010, SchleicherDreizler2014, Veras2016}.\\
Another possibility is the existence of a hybrid  generation: first generation planets that survive the first CE phase and may have been subject to mass loss throughout the whole process. Either way, the resulting planet/planetesimal could now accrete on the disc material, producing more massive planets on higher eccentricity \citep{ArmitageHansen1999, Perets2010}.
The outcome would be a planet with a first generation inner core and second generation outer layers. In this case the formation of a giant planet could be faster than for first generation giant planets.

The same hypothesis is similarly applicable if the binary overgoes a second CE phase, i.e. the low-mass star overflows its Roche lobe and shares its atmosphere with the WD companion. After this stage we might have either a third generation of exoplanets forming around a double white dwarf (DWD) system, a hybrid generation, or previous surviving generations. 
To date no exoplanets are known orbiting a DWD \citep{TamaniniDanielski2019}, and the only circumbinary exoplanet known orbiting a system with two post-main sequence stars (i.e. a WD and a millisecond pulsar) is the giant PSR B1620-26AB b;  this planet is also the first circumbinary
exoplanet confirmed  \citep{Sigurdsson1993, Thorsett1993}. Because the planetary system PSR B1620-26AB is the result of a stellar encounter in the Milky Way plane \citep{Sigurdsson2003}, it is not directly representative of a standard (i.e. isolated) binary planetary system evolution.

Possibly because of an observational bias, all the post-CE planets discovered until now are giant planets with masses $M \geq$ 2.3 $\Mj$ and semi-major axes $a \geq$ 2.8 au. The most successful technique used for their detection is eclipse timing variation (ETV), which is sensitive to wider planetary orbits and hence requires a long observational baseline to precisely time the eclipses.
Also, ETV typically suffers from a lack of cross validation and errors that are not uncommon, 
for example, the lack of accurate timing in the instrumentation used or procedures used to place the recorded times onto a uniform timescale corrected for light travel time \citep{Marsh2018}. Any small inaccuracy or analysis imprecision could lead to uncertainties in the validity of a
planet in the system, with the planets potentially being the wrong interpretation of the Applegate mechanism \citep{Applegate1992}.

Recently \citealt{TamaniniDanielski2019} showed the possibility to detect \textit{Magrathea}-like \citep{Magrathea} planets, 
i.e. circumbinary exoplanets orbiting DWDs by using the Laser Interferometer Space Antenna (LISA; \citealt{LISAcallpaper}) to measure the characteristic periodic modulation in the gravitational wave (hereafter GW) signal produced by the DWD.
Compared to the classic detection methods, the GW approach has the advantages that this method is (i) able to find exoplanets all over the Milky Way and in other close-by galaxies; (ii) not limited by the magnitude of the WDs, but on the parameters from which the GW depends (see Section \ref{sec:LISAdetection}); and (iii) not affected by stellar activity, which is an issue reported in electromagnetic (EM) observations.

\subsection{Brown dwarfs}
Brown dwarfs (hereafter BDs) are by definition bodies that are not massive enough to fuse hydrogen in their interior stably,
but are massive enough to undergo a brief phase of deuterium burning soon after their formation.
The very first two BDs were discovered in 1995 \citep{Nakajima1995, Rebolo1995}, and today over 1000 BDs have been detected in the solar neighbourhood \citep{Burningham2018}. 
Some of these objects have also been discovered around single WDs. Examples of BDs orbiting at distances beyond the tidal radius of the asymptotic giant branch progenitor, but also within it (e.g.  WD 0137-349 B, \citealt{Maxted2006}), show that BDs can survive stellar evolution whether or not they are engulfed by their host's envelope. \cite{Farihi2005} predicted that a few tenths of percent of Milky Way single WDs host a BD.\\
Concerning the binary case, the ETV technique allowed observers to detect 
a few post-CE systems with one evolved binary, comprised of a WD and a low-mass star, and a BD companion(s). 
Some examples of such systems are HQ Aqr, V471 Tau, HW Vir, and KIC 10544976  \citep{Gozwski2015, Vaccaro2015, Beuermann2012B, Almeida2019}.
No BD has been found orbiting DWDs but, similar to the case of circumbinary planets (CBPs), if such a population exists, it could be found 
through GW astronomy with the LISA mission \citep{Robson2018, TamaniniDanielski2019}. 
As a matter of fact a BD, because it is more massive than a planet, would produce a stronger GW perturbation 
that is easier to detect with respect to a CBP. 

Recalling the hypothesis of an hybrid generation (Sec \ref{sec:postCEexoplanets}),
the core of a surviving body could efficiently accrete on the stellar ejecta disc, forming exceptionally 
massive planets which de facto become BDs \citep{Perets2010}.
In this case BDs would be able to form more often within the famous BD desert \citep{MarcyButler2000}.

\subsection{Detecting sub-stellar objects around binary WDs with LISA}
\label{sec:outline}

The focus of this work is to follow up on \cite{TamaniniDanielski2019} and to quantitatively estimate the LISA detection rates of circumbinary exoplanets as well as circumbinary BDs.
Brown dwarfs have masses ranging between the stellar and planetary domain; nevertheless, while the difference with stars is well defined, the separation with planets is still an open subject of discussion. The different nature of these objects could be either based on their intrinsic physical properties \citep{ChabrierBaraffe2000} or on their formation mechanism \citep{Whitworth2007}.
Furthermore, more recently \citet{HatzesRauer2015} analysed the density versus mass relationship for objects with mass 
$\sim$ 0.01 $\Mj$ < $M$ < 0.08 $\Msun$, and identified three distinct regions that are separated by 
a change in slope in such a relation (at $M$ = 0.3 $\Mj$ and $M$ = 60 $\Mj$). 
Above $M$ = 60 $\Mj$, but lower than $M$ = 0.08 ~$\Msun$, the BDs domain, and below that limit 
(but above $M$ = 0.3 $\Mj$) the  giant planets domain.

Because of this ongoing discussion we hence decided to not limit our analysis to the mass domain reported in \cite{TamaniniDanielski2019}, but to account for a larger mass range, up to the stellar limit.
Consequently throughout this manuscript, for simplicity we define a sub-stellar object (hereafter SSO) to be a celestial body with mass less than 0.08 $\Msun$ (the hydrogen burning limit, which includes the upper uncertainty by \citealt{Whitworth2018}). 
This category is divided between CBPs and BDs. 
As in \cite{TamaniniDanielski2019} we define the former 
as objects with mass $M \leq$ 13 $\Mj$ (the deuterium burning limit) and the latter as those with mass 13 $\Mj < M < 0.08 ~\Msun$. 
For simplification only the mass and no spectroscopic and/or formation mechanism classification are accounted for in this work.

The outline of this manuscript is as follows: 
in Section \ref{sec:method} we present the characteristics of populations used in the investigation, and we
summarise the 
GW detection method discussed in \citealt{TamaniniDanielski2019}. 
In Section \ref{sec:results} we report CBPs and BDs detection rates, with their error analyses,  for both the LISA nominal mission
and for a possible extension of four more years.
We discuss the implications of our results in Section \ref{sec:discussion} and we conclude in Section \ref{sec:conclusions}.

\section{Method}
\label{sec:method}

To reach the scope of this study we worked throughout two different stages.
First, we constructed a population of Galactic detached
DWDs with circumbinary exoplanets/BDs. To do so we injected a simulated population of SSOs into a synthetic population of DWDs \citep{Korol2017}. 
Such a DWD population was specifically designed to study the LISA detectability of these binary WDs, and it was employed in the LISA mission proposal \citep{LISAcallpaper}.  
Second, we used the method described in \citet{TamaniniDanielski2019} to measure how many SSOs LISA will be able to detect. 

In Section \ref{sec:dwds} we summarise the most important features of the DWDs population. 
In Section \ref{sec:injection} we provide details about the SSOs population injection process.
In Section \ref{sec:LISAdetection} we summarise the method used for the LISA GW detection of a circumbinary SSOs.

\subsection{LISA DWD population} \label{sec:dwds}

\begin{figure*}[t]
    \centering
    \includegraphics[width=.8\textwidth]{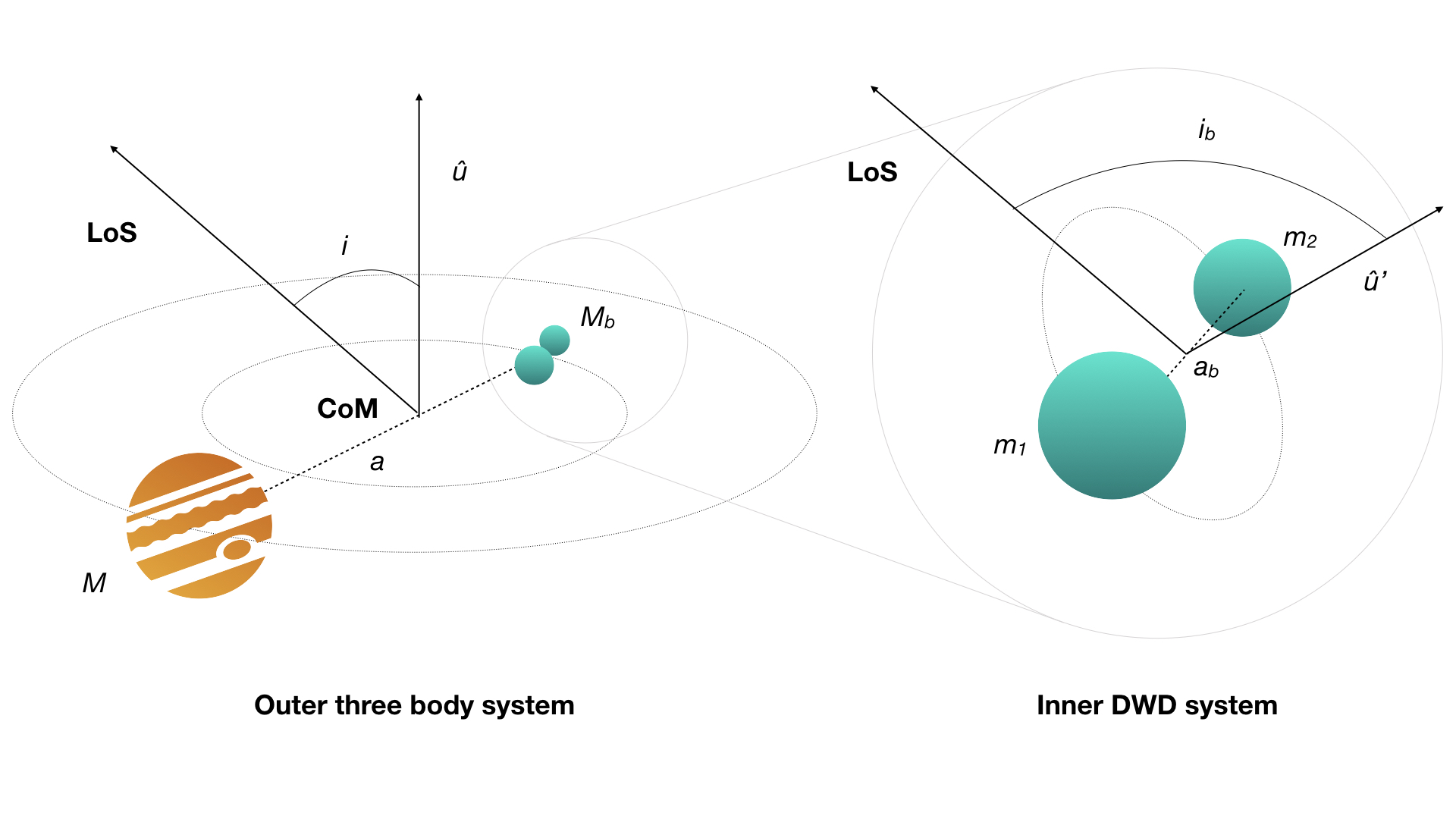}
    \caption{
    Geometry of the outer three-body system (DWD+planet/BD) and inner compact two-body system (DWD).
    The quantities $\hat{u}$ and $\hat{u}'$ denote the directions perpendicular to the outer and inner orbital planes, respectively.
    The acronyms LoS and CoM stand instead for line of sight and centre of mass (of the whole three-body system).
    }
    \label{fig:geometry}
\end{figure*}

\begin{figure*}[t]
    \centering
         \includegraphics[width=0.49\textwidth]{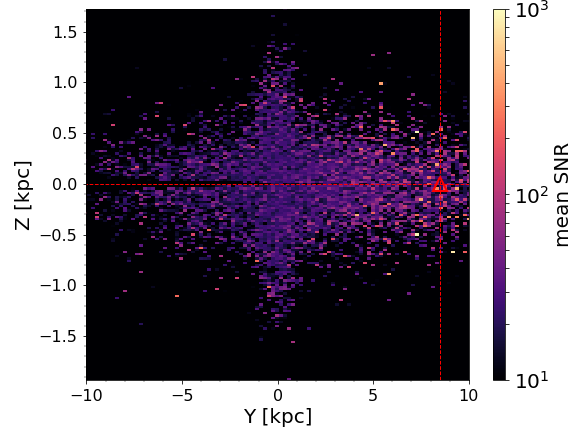}
         \includegraphics[width=0.49\textwidth]{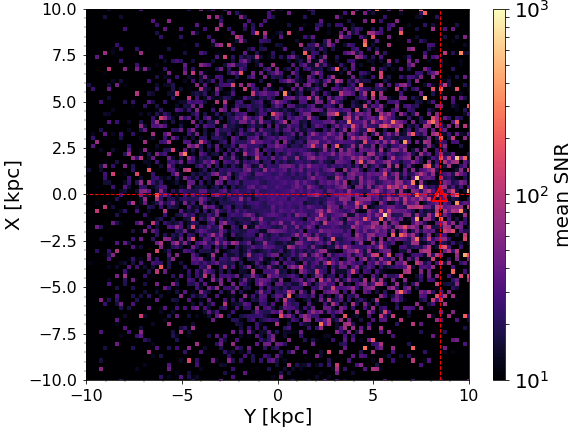}
         \caption{Signal-to-noise map of DWDs detected by LISA (4 yr) in the galactocentric Cartesian coordinate system. The colour represents the mean S/N per bin. The red triangle identifies the position of the LISA detector in our simulation.}
       \label{fig:MWinGWs}
\end{figure*}

Our method relies on the binary population model which \cite{too12} obtained using binary population synthesis code {\sc SeBa}, originally developed by \citet{por96} and later adapted for DWDs by \citet{nel01} and \citet{too12}. 
The progenitor population is initialised by randomly sampling initial distributions of binary properties with a Monte Carlo technique.
Specifically, the mass of the primary star is drawn from the Kroupa initial mass function in the range between 0.95 and 10$\,$M$_{\odot}$ \citep{Kroupa1993}. 
The mass of the secondary star is derived from a uniform mass ratio distribution between 0 and 1 \citep{Kraus2013}. 
A log-flat distribution and a thermal distribution are adopted for the initial binary orbital separations and binary eccentricities, respectively \citep{Abt1983,Heggie1975,Kraus2013}. 
The initial binary fraction is fixed to 0.5 value. The 
{\sc SeBa} code evolves binaries until both stars turn into WDs and beyond up to the present time. More details and discussion on the sensitivity of the binary population synthesis outcome to the aforementioned assumptions are given in \citet{too12,Toonen17}. The adopted model has also been recently tested against observations of both single WDs and WDs in binary systems (including DWDs) in the solar neighbourhood by \cite{Toonen17}. 
In particular, the adopted model currently better represents the space density of DWDs derived from a spectroscopically selected sample of \citet{Maoz18}.

One of the most impacting assumptions in DWD population synthesis is the prescription for the CE evolution \cite[e.g.][]{Toonen17}.
As mentioned in Section ~\ref{sec:intro}, CE is a short phase of the binary evolution in which the more massive star of the pair expands and engulfs its companion \citep{Paczynski1976,Webbink1984}. 
During the CE phase the binary orbital energy and angular momentum can be transferred to the envelope because of the dynamical friction that the companion star experiences when moving through the envelope. 
Typically, this process is implemented in the binary population synthesis either by parametrising the conservation equation for energy (through the $\alpha$ parameter) or that for angular momentum (through the $\gamma$ parameter) \citep[see][for a review]{Ivanova2013}.
In particular, the $\gamma$-prescription was introduced with the aim to reconstruct the evolution path of observed DWDs by \cite{nel00,Nelemans2005}. 
In the model adopted for this study, $\gamma \alpha$, we allowed both parametrisations; the $\gamma$-prescription was applied unless the binary contains a compact object or the CE is triggered by a tidal instability. 
It has also been shown that $\gamma \alpha$ model describes observations better than the model in which only $\alpha$-prescription is employed \citep{too12}. 
Future optical surveys such as the Large Synoptic Survey Telescope \citep[LSST;][]{LSST} will provide large samples of new DWDs that will help to further constrain CE evolution for these systems \citep{Korol2017}.

Next, we distributed DWDs in a Milky Way-like galaxy according to a star formation history.
We adopted a simplified Galactic potential composed of an exponential stellar disc and a spherical central bulge.
Similarly to \cite{Ruiter09,lam19}, we found that the contribution of the stellar halo to the total amount of detectable GW sources is at most of a few percent. Thus, it is not included in this study. 
We populated the disc according to the star formation rate (SFR) from \citet{boi99} and assumed the current age of the Galaxy to be 13.5 Gyr. 
To model the bulge of the Milky Way we doubled the SFR in the inner $3\,$kpc as in \citet{nel04}. 
The detailed description of the Galactic model is presented in \cite{Korol2019}.
Finally, we assigned binary inclination angle $i_{\rm b}$, drawn from a uniform distribution in $\cos i_b$.
Thus, each DWD in the catalogue is characterised by seven parameters: $m_1,\ m_2,\ P_b,\ i_b$, the Galactic latitude $l$ and longitude $b$, and the distance from Sun $d$ (see Figure \ref{fig:geometry}).

To obtain a sub-sample of DWDs detectable by LISA we employed the Mock LISA Data Challenge (MLDC) pipeline, designed for the analysis of a large number of GW sources simultaneously present in the data \cite[e.g. ][]{Littenberg2013,Cornish17}. 
This is realised throughout an iterative process that is based on a median smoothing of the power spectrum of the input population to compute the overall noise level (instrument plus confusion from the input population). 
The resolved sources (i.e.~those with S/N $ > 7$) are extracted from the data until the convergence. 
We adopted the LISA noise curves and orbits according to the latest mission design, the nominal mission duration of four years and the extended mission duration of eight years \citep{LISAcallpaper}.

We find approximately $26 \times 10^3$ and $40 \times 10^3$ detached DWDs with S/N > 7 for the nominal four years and extended eight years of the LISA mission duration, respectively. 
Figure~\ref{fig:MWinGWs} illustrates the distribution of detected DWD in our mock  Galaxy showing that GW detections can map both disc and bulge at all latitudes.
We represent the mean S/N per bin in colour.

We note that in this work we focus on detached DWD binaries only. In principle, other Galactic binaries composed of compact objects (such as WD - neutron star and double neutron stars) and accreting systems could also host a CBP/BD. However, these are significantly less abundant in the Milky Way \citep[e.g.][]{nel04}, and thus would not affect much our estimates. In addition, GW signal of accreting systems contains an imprint of the mass-transfer process, which could affect the detection of circumbinary companions. We leave these investigations for future work.

\subsection{Exoplanet and brown dwarf injection}
\label{sec:injection}

Since the WD pollution effect supports evidence of dynamically active planetary
systems around single WDs (Section \ref{sec:wdpollution}) and since no data are available for the binary WD case, we set the WD pollution upper limit occurrence rate (i.e.  50$\%$ \citealt{Koester2014}) to be the occurrence rate (hereafter O.R.) of the synthetic population of SSOs orbiting DWD. We neglected the presence of an external third star and we assumed that pollution derives from asteroidal or moon material, rather than cometary material. We also rejected exceptions such as the capture of a free-floating planet at thousands of astronomical units, and we assumed that each DWD can harbour only one SSO; we briefly discuss the implications of considering multiple circumbinary objects in Sec.~\ref{sec:discussion}. 

For the following we note that co-evolution of the binary plus SSO was neglected, and that the SSO population was injected into already formed WD-WD systems in which the stability criterion ($P\ \gtrsim 4.5 \ P_b$) of \cite{HolmanWiegert1999} was always satisfied.

In accordance with the pollution O.R. employed in this investigation, we set the SSO maximum distance ($a$) to be 
the approximate maximum limit for pollution to occur.
Given that the maximum distance at which those asteroids reside around DWDs is completely unconstrained \citep{Veras2019}, we assumed 200 au to be a reasonable distance at which the SSO could still perturb asteroids which lie outwards or inwards towards the binary.

We set a uniform SSO inclination  in $\cos i$ (cf.~Figure  \ref{fig:geometry}) and uniform initial phase $\phi_0$ 
between 0 -  2$\pi$.
Given that the planet distribution function is unknown and that no compelling physical motivation for a specific model at wide separations exists for these systems, we tested a combination of various semi-major axis $a$ and SSO mass $M$ distributions, commonly  presented in the literature,  to measure the number of possible detection of both CBPs and BDs. 
More specifically we defined the semi-major axis distributions as follows:

A) uniform distribution $\mathcal{U}_a$ (0.1 - 200 au); and 
B) $\log{}_{10}$ uniform distribution $\log{\mathcal{U}_a}$ (0.1 - 200 au); and
C) log-normal distribution $f(x) = A\, e^{\ln(x)-\mu / 2\sigma^2} / (x 2\pi \sigma),$ where  $A$ is the amplitude, $\mu$ the mean of the log-normal, and $\sigma$ the square root of the variance.  \cite{Meyer2018} give more details and specific values of the parameters; and
D) power-law distribution a$^{-0.61}$ (0.1 - 200 au) \citep{Galicher2016}.

The mass distributions are as follows:
1) uniform distribution $\mathcal{U}_M$(1 $\Mearth$ - 0.08 $\Msun$); and
2) a combination of power law, $M^{-1.31}$  \citep{Galicher2016} between 1$\Mearth$ - 13$\Mj$ and uniform distribution for 13 $\Mj$ < $M$ < 0.08 $\Msun$.


\subsection{LISA detection of a third sub-stellar object}
\label{sec:LISAdetection}

\begin{figure}[t]
    \centering
    \includegraphics[width=\columnwidth]{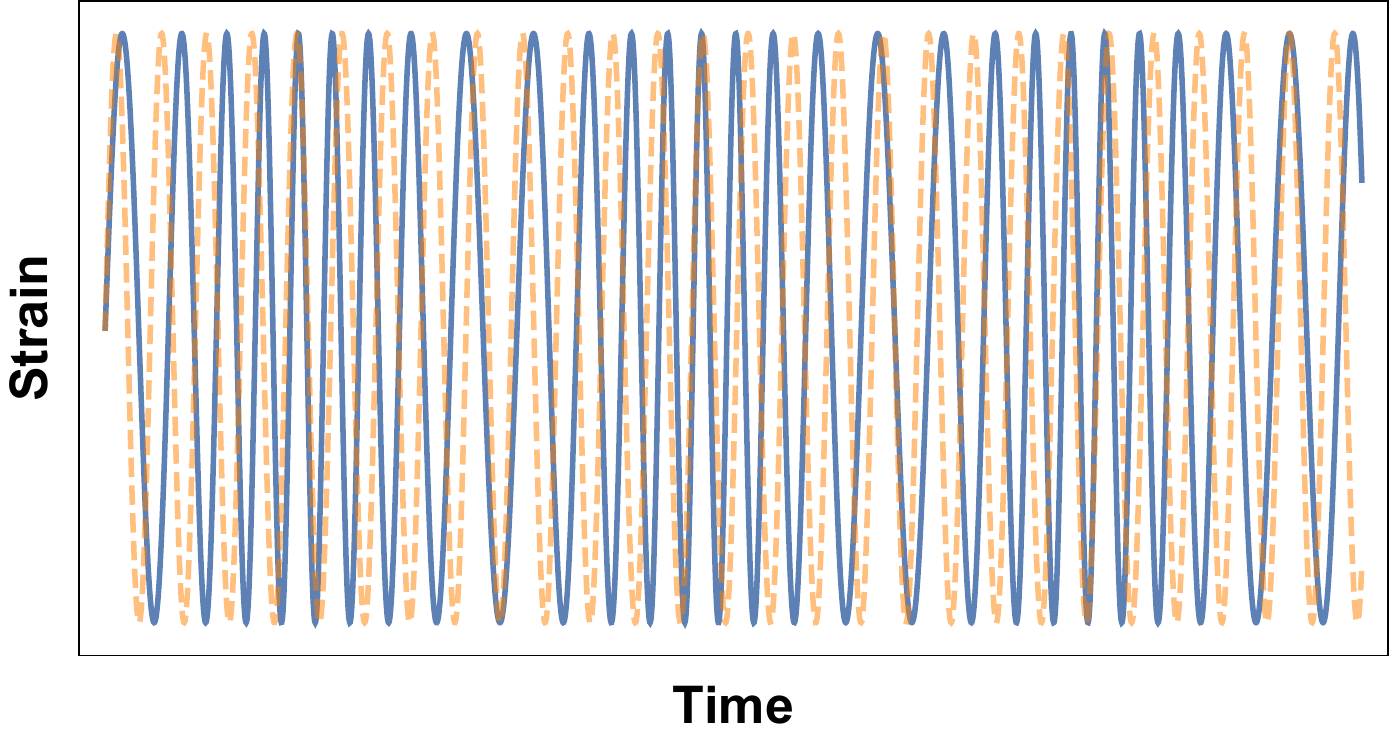}
    \caption{
    Qualitative example of a DWD waveform with (blue) and without (orange dashed) the presence of a third body. The Doppler modulation is extremely exaggerated for visualisation purposes.
    }
    \label{fig:waveform_example}
\end{figure}

To model the perturbation induced by the SSO on the GW signal emitted by the DWDs, we followed the procedure presented in \cite{TamaniniDanielski2019}. Figure \ref{fig:geometry} shows the geometry of the three-body system under consideration.
The motion of the DWD around the centre of mass of the three-body system modulates the GW frequency through the well-known Doppler effect.
The resulting frequency observed by LISA is written as\begin{equation}
 f_{\rm obs}(t) = \left( 1 + \frac{v_{\parallel}(t)}{c} \right) f_{\rm GW}(t) \,,
 \label{eq:Doppler}
\end{equation}
where $v_\parallel$ is the line-of-sight velocity of the DWD with respect to the common centre of mass, while $f_{\rm GW}$ is the GW frequency in the reference frame at rest with respect to the DWD centre of mass.
Since the DWDs observed by LISA do not merge before a time much larger than the observational lifetime of the mission, we can effectively model the emitted frequency with a Taylor expansion around a constant value and only keep the first order term
\begin{equation}
 f_{\rm GW}(t) = f_0 + f_1 t + \mathcal{O}(t^2) \,,
\end{equation}
where $f_0$ is the frequency when LISA starts taking data and $f_1$ is its first derivative evaluated at the same time.
\afterpage{%
\noindent
\centering
\begin{minipage}{\textwidth}
\centering
\resizebox{17cm}{!}{%
    \renewcommand{\arraystretch}{1.5} 
    \begin{tabular}{c|c c|c c|c c|c c }
    \multicolumn{9}{c}{DETECTIONS (4 years)}\\
    \hline\hline
     &  \multicolumn{2}{c|}{A)  $\mathcal{U}_a$ (0.1-200 au)} &  \multicolumn{2}{c|}{B)  $\log{\mathcal{U}_a}$ (0.1-200 au)} &  \multicolumn{2}{c|}{C)  logNormal$_a$ (0.1-200 au)} &  \multicolumn{2}{c}{D)  $a^{-0.61}$ (0.1-200 au)}  \\
     \cline{2-9}
     & CBPs & BDs & CBPs & BDs & CBPs & BDs & CBPs & BDs \\
     \hline
    1) $\mathcal{U}_M$ (1$\Mearth$ - 0.08 $\Msun$) &  {\bf 3 (0.011\%)}  & 79 (0.302\%) & {\bf 83 (0.317\%)} & {\bf 2218 (8.482\%)} & 18 
    (0.069\%)  & 503  (1.924\%)  & 28 
    (0.107\%) & 820 (3.136\%) \\
    \hline
    2) $M^{-1.31}$ & 6 (0.023\%) & {\bf 14 (0.054\%)} & 30 (0.115\%) & 316 
    (1.209\%) & 5 (0.019\%) & 85 (0.325\%) & 13 (0.050\%) & 131 (0.501\%)\\
    \hline
    \end{tabular}
    }
    \captionof{table}{Number of planetary detections depending by the different combinations of mass and semi-major axes distributions. In bold the minimal and maximal values for both CBPs and BDs. The percentage is computed over a total of 26148 DWDs (visible during the nominal LISA mission length).}
 \label{tab:detections}
 
\vspace{0.5cm}
\centering
      \includegraphics[trim=0.1cm 0 0.3cm 0.2cm, clip,width=0.45\linewidth]{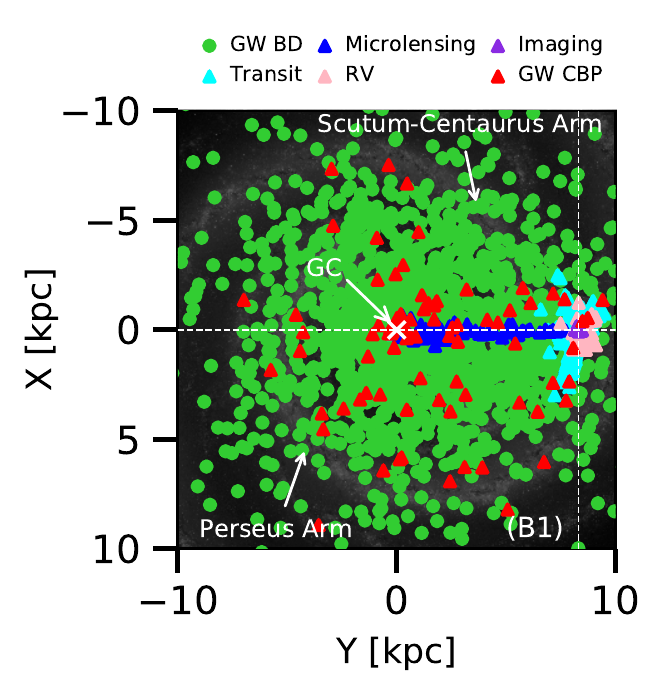} \quad
      \includegraphics[trim=0.1cm 0 0.3cm 0.2cm,
      clip,width=0.43\linewidth]{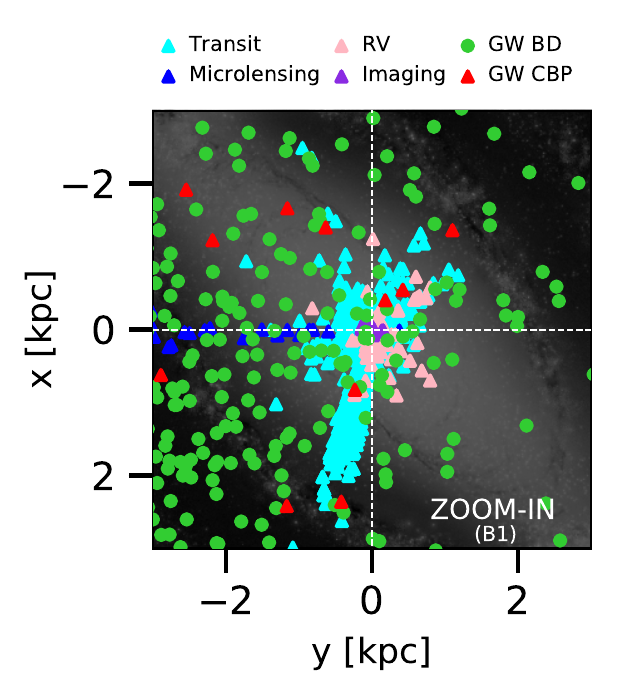}\\
     
      \includegraphics[trim=0.1cm 0 0.3cm 0.2cm, clip,width=0.45\linewidth]{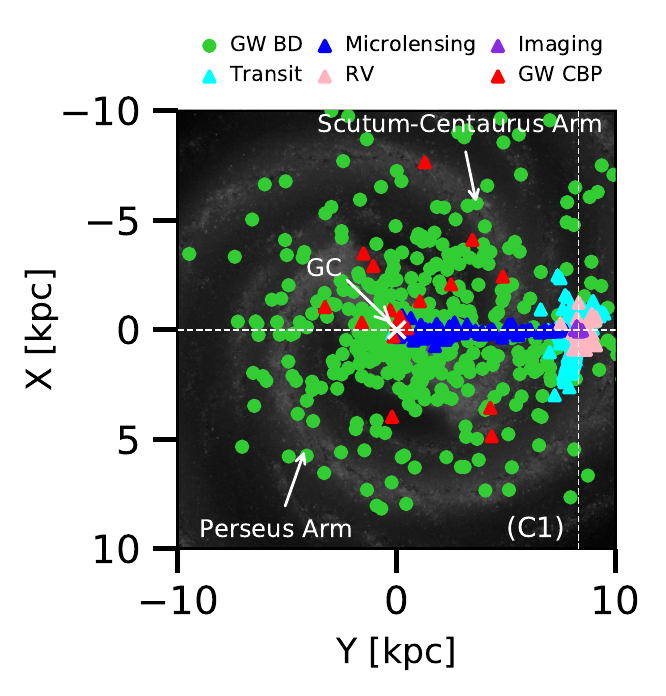}\quad
      \includegraphics[trim=0.1cm 0 0.3cm 0.2cm,
      clip,width=0.45\linewidth]{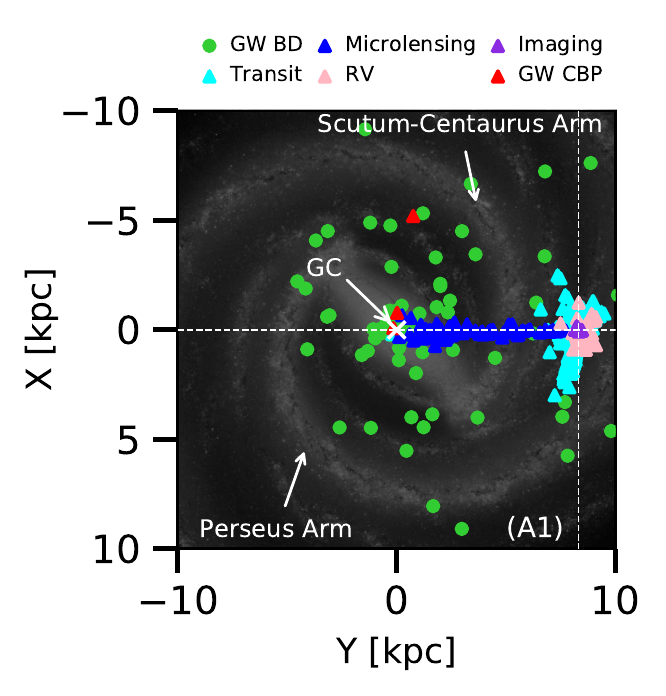}
    \captionof{figure}{Optimistic (top left,  B1) with its zoom-in on the solar region (top right, heliocentric coordinates), intermediate  (bottom left, C1), and pessimistic (bottom right, A1) scenarios. Each plot shows the location of the binary WD system with a planetary companion (red) and BD (green) detection through GWs.
    In each panel we also plot the known detected exoplanets's host-star (see legend for colour scheme; data from \url{https://exoplanetarchive.ipac.caltech.edu}). We note that data overlay a face-on black and white image of the Milky Way for Galactic location reference purposes.
    }
    \label{fig:detectionsGalaxy}
\end{minipage}
\clearpage  %
}%
The line-of-sight velocity of the DWDs is instead given by
\begin{equation}
 v_\parallel = - K \cos\varphi(t) \,,
\end{equation}
where we defined the parameters
\begin{equation}
 K = \left(\frac{2\pi G}{P}\right)^{\frac{1}{3}} \frac{M}{(\Mb+M)^{\frac{2}{3}}} \sin i  \,,
 \label{eq:K}
\end{equation}
and the orbital phase 
\begin{equation}
    \varphi(t) = \frac{2\pi t}{P} + \varphi_0 \,,
\end{equation}
both derived assuming an SSO circular orbit.
In the expressions above $P$ is the SSO orbital period, $M$ is the SSO mass, $\Mb$ is the DWD total mass, $\varphi_0$ is the outer orbital initial phase, and $i$ is the SSO orbital inclination (cf.~Figure \ref{fig:geometry}).
The phase of the waveform observed by LISA is then given by
\begin{equation}
 \Psi_{\rm obs}(t) = 2\pi \int f_{\rm obs}(t') dt' + \Psi_0 \,,
\end{equation}
where $\Psi_0$ is a constant initial phase.
The main contribution of the Doppler frequency modulation \eqref{eq:Doppler} consists in a periodical shift of the GW frequency towards higher and lower values around $f_0$.
This effect is qualitatively depicted in Figure  \ref{fig:waveform_example} in which the Doppler modulation has been extremely exaggerated with respect to the perturbation induced by a SSO on a DWDs. In the real case the modulation timescale, of the order of  roughly years, is much longer than the period of the GW produced by the binary, roughly minutes, implying that the effect would not be visible by eye.

For each DWDs in our mock catalogue we can thus build a waveform depending on 11 parameters: 8 parameters associated with the DWD, namely $\ln(A), \Psi_0, f_0, f_1, \theta_S, \phi_S, \theta_L, \phi_L$, and 3 parameters associated with the SSO orbit, namely $ K, P, \varphi_0 $. In this case $\theta_S, \phi_S, \theta_L, \phi_L$ are the two sky localisation angles and the two angles defining the orbital geometry of the DWDs, respectively (directly related to the inclination $i_b$ and polarisation angle $\psi_b$; see e.g.  \citet{Cornish:2003vj}).

To simulate the response of LISA and perform a parameter estimation of the GW waveform, we followed \cite{TamaniniDanielski2019} again.
The full expressions for the two linearly independent signals observed by LISA $h_{I,II}(t)$, including the LISA antenna pattern functions and effects due to its orbital motion, can be found in \cite{Cutler1998,Takahashi:2002ky,Cornish:2003vj}.
For the sake of simplicity we are not reporting those expressions in this work. 
The S/N of each event is computed as the following:
\begin{equation}
 {\rm S/N}^2 = \frac{2}{S_n(f_0)} \sum_{\alpha = I,II} \int_0^{T_{\rm obs}} dt\, h_\alpha(t) h_\alpha(t) \,,
 \label{eq:SNR}
\end{equation}
where $T_{\rm obs}$ is LISA observational time period and $S_n(f_0)$ is the one-sided spectral density noise of LISA computed at $f_0$.
Parameter estimation is performed by employing a Fisher information approach, where we define the Fisher matrix as
\begin{equation}
 \Gamma_{ij} = \frac{2}{S_n(f_0)} \sum_{\alpha = I,II} \int_0^{T_{obs}} dt \frac{\partial h_\alpha(t)}{\partial\lambda_i}\frac{\partial h_\alpha(t)}{\partial\lambda_i} \,.
 \label{eq:FM}
\end{equation}

\noindent Marginalised 1$\sigma$ errors for each waveform parameter are thus estimated from the square root of the diagonal elements of the covariance matrix, the inverse of the Fisher matrix.


\section{Results}
\label{sec:results}
We focus first on the properties of the detected population of SSOs (Sec.~\ref{subsec:detections}), showing also how the numbers improve for an extended eight-year LISA mission (Sec.~\ref{subsec:detections_8yrs}). We then discuss the recovered accuracy on the waveform parameters in Sec.~\ref{sec:errordistribution}.

\subsection{LISA detection of SSOs}
\label{subsec:detections}

As in \cite{TamaniniDanielski2019} we assume that a SSO (either a CBP or a BD) is detected if both $K$ and $P$ parameters are measured with a relative accuracy better than 30\%.
For every injected SSO population, defined by a combination of semi-major axis $a$ and mass $M$ distributions (see Sect. \ref{sec:injection}), we counted the number of SSOs whose GW perturbation can be detected by LISA. 

We report in Table \ref{tab:detections} the total number and percentage of circumbinary exoplanets and BDs detected during the nominal LISA mission length. 
For both CBPs and BDs we identified optimistic, pessimistic, and intermediate scenarios. While the first and second represent the cases in which the highest and lowest numbers of CBPs (or BDs) are detected, the last scenario represents the case with the median number of detections, rounded by excess.
Among the available combinations, the B1 scenario, i.e. that whose injected SSO population follows a logarithmic $a$ distribution $\log{\mathcal{U}_a}$, and uniform $M$ distribution $\mathcal{U}_M$ (see Sect.~\ref{sec:injection}), is the optimistic case for both CBPs and BDs with 83 and 2218 detections, respectively.
The intermediate scenario is represented by C1 (log-normal$_a$; $\mathcal{U}_M$), and B2 ($\log{\mathcal{U}_a}$; $M^{-1.31}$), for CBPs and BDs with 18 and 316 detections, respectively. 
The pessimistic scenario is represented by A1 ($\mathcal{U}_a$;$\mathcal{U}_M$) and A2 ($\mathcal{U}_a$; $M^{-1.31}$) with 3 and 14 detections for CBPs and BDs, respectively.
We plot in Figure \ref{fig:detectionsGalaxy} the location in the Milky Way of the detections for the three CBPs scenarios together with a zoom-in on the solar neighbourhood for the optimistic scenario.
From Figure \ref{fig:detectionsGalaxy} it is easy to understand that LISA will be able to observe CBPs and BDs orbiting DWDs everywhere in the Galaxy.

Furthermore, for the six scenarios selected above, Figure \ref{fig:CBP_distributions_log} and Figure \ref{fig:BD_distributions_log} (currently appearing after the references) show the distribution of detected CBPs and BDs, respectively, over the CBP/BD separation from the DWD ($a$), the mass of the CBP/BD ($M$), the CBP/BD orbital inclination ($i$), the parameter $K$, the CBP/BD period ($P$), the DWD period ($P_b$), the DWD S/N, the distance from the Earth ($d$), the DWD chirp mass ($M_c$), and the total DWD magnitude measured in the Gaia $G$ band ($G_{\rm DWD}$).
To highlight possible observational biases, in Figs.~\ref{fig:CBP_distributions_log} and \ref{fig:BD_distributions_log} we also show the underlying distribution of injected CBPs/BDs in grey.

\subsection{Detection rates for an extended LISA mission}
\label{subsec:detections_8yrs}

\begin{SCtable*}[0.9][t]
    \renewcommand{\arraystretch}{1.5} 
    \centering
    \begin{tabular}{c|c c}
        \multicolumn{3}{c}{DETECTIONS (8 years)}\\
        \hline\hline
         & CBPs & BDs  \\
        \hline
        B1 & 215 (0.822\%) & 4684 (17.913\%) \\
        \hline
        \hline
        A1 & 8 (0.02\%)  & 295 (0.733\%)\\
        \hline
        A2 & 11 (0.027\%) & 43 (0.107\%) \\
        \hline
        \end{tabular}
        \hspace{0.4cm}
    \caption{
    Number of detections (over a total of 40251 DWDs detected with 8 yr LISA mission) for the best and worst scenarios for both CBPs (B1 and A1) and BDs (B1 and A2).}
    \captionsetup[SCtable][c]{justification= RaggedRight}.
    \label{tab:8yrsdetections} 
\end{SCtable*}

\begin{figure*}[h!]
\vspace{0.5cm}
    \centering
    \includegraphics[width=0.83\textwidth]{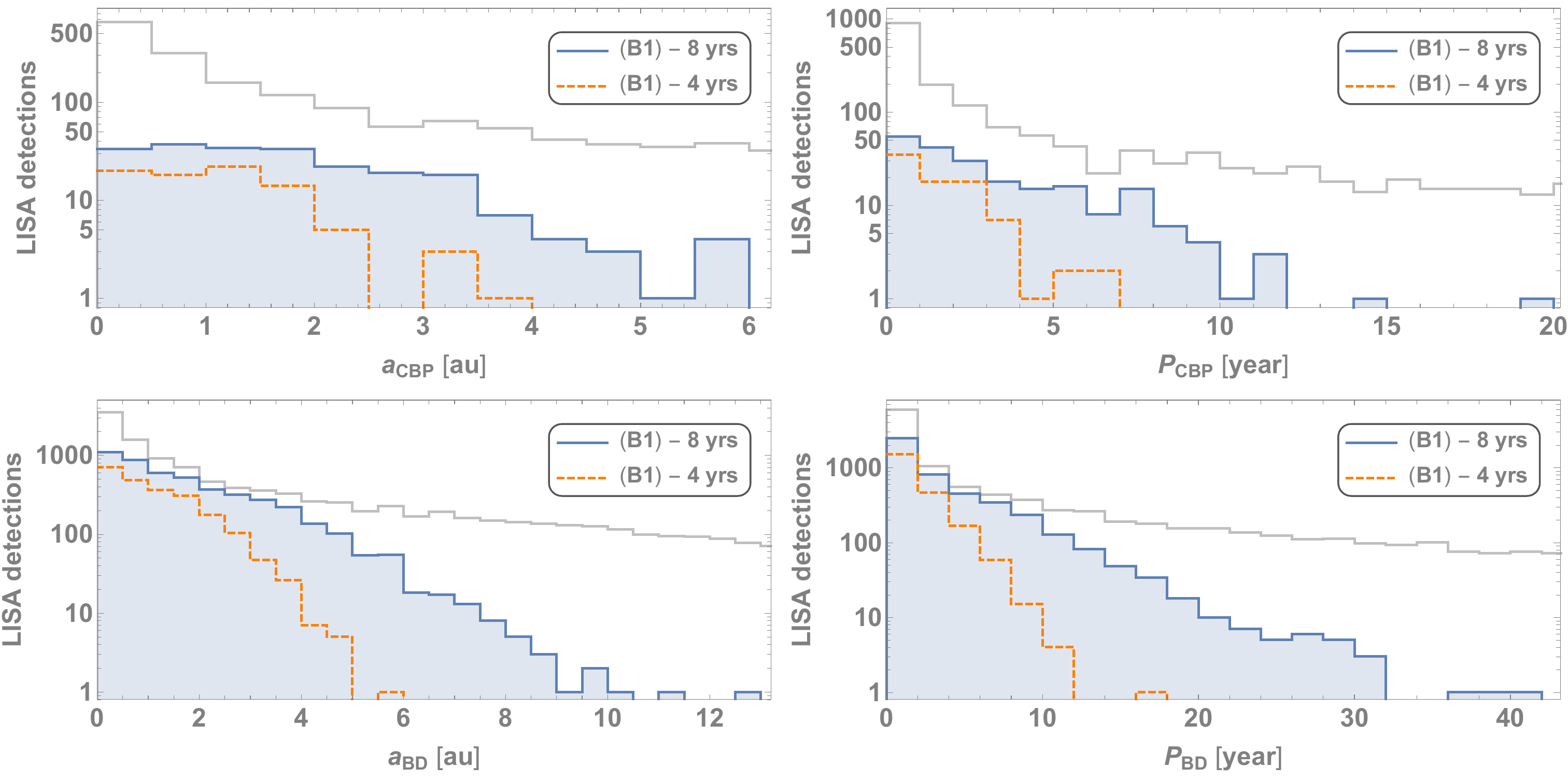}
    \captionof{figure}{Comparison between injected (grey) vs. detected population distributions ($a$; $P$) 
    for 8 yr (blue) observations versus the 4 yr detected population (dotted orange).
    Only the optimistic scenario (B1) is shown for both CBPs (top panels) and BDs (bottom panels).
    }
    \label{fig:8yrs_distributions}
\end{figure*}

\begin{figure*}[h!]
    \centering
    \includegraphics[width=0.82\textwidth]{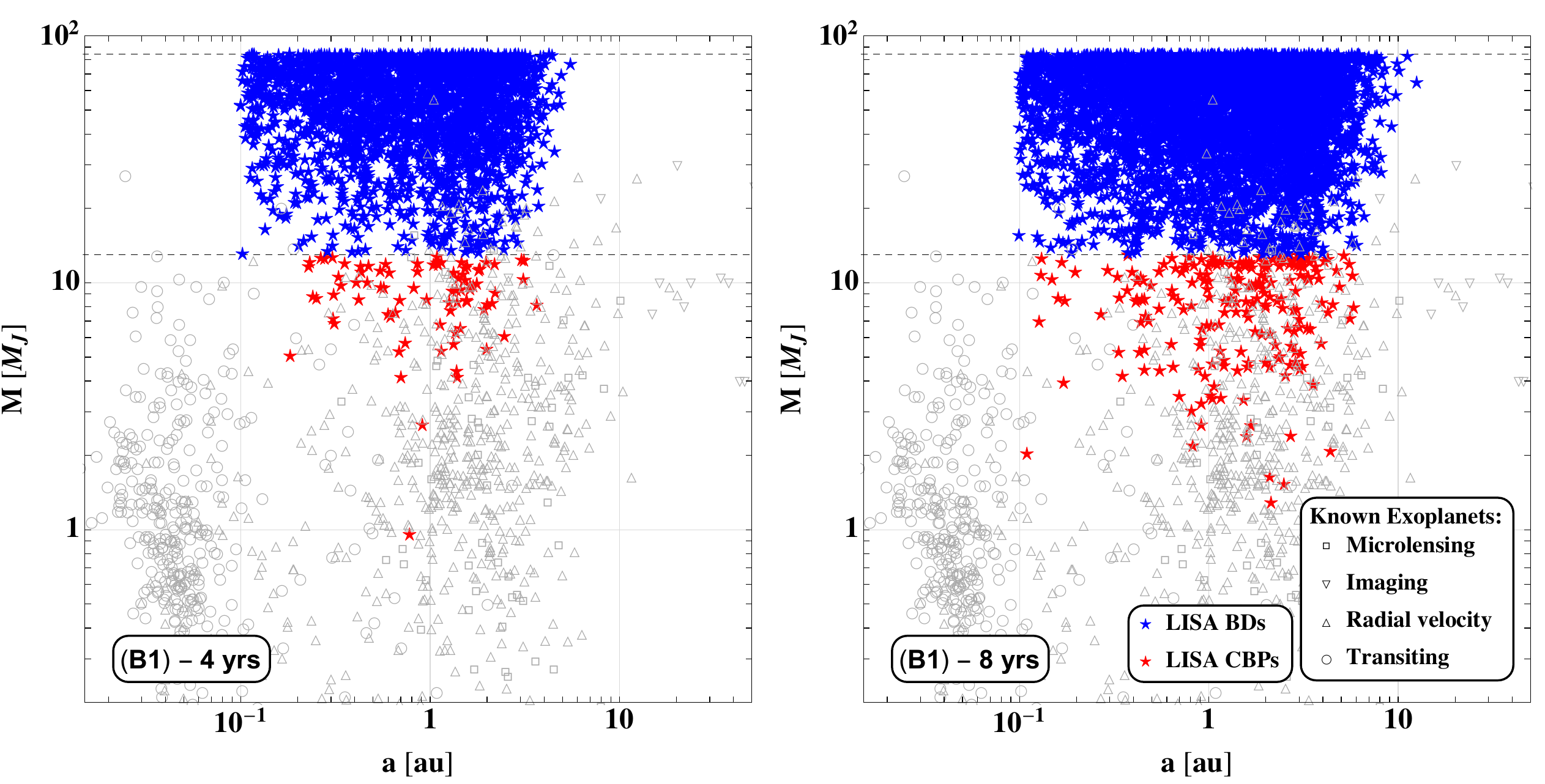}
    \captionof{figure}{Detections by LISA\ of circumbinary exoplanets (red) and BDs (blue) in the optimistic (B1) scenario for 4 yr (left panel) and 8 yr (right panel) of observations. The SSOs mass ($M$, or $M \sin(i)$ for those planet whose inclination is not known yet), as a function of the SSO-to-binary separation ($a$) is shown, together with values of known exoplanets (data were taken from \url{https://exoplanetarchive.ipac.caltech.edu}).
    The horizontal dotted line corresponds to the limits in mass we explored: 13 $\Mj$ and 0.08 $\Mj$.
    }
    \label{fig:sepVmass}
\end{figure*}
We repeated our analysis for an eight-year LISA mission, corresponding to a possible realistic extension beyond the nominal four-year mission; this can also approximately be considered as ten years of mission operations, the maximal envisaged extended duration, with duty cycle of 80\% similar to the LISA Pathfinder ( \citealt{LPF2016}).
We used the catalogue of $40 \times 10^3$ DWD detected over the eight years of mission presented in Section \ref{sec:dwds} injecting SSOs according to the optimistic and pessimistic scenarios only.
The total detections of CBPs and BDs, together with the percentage over the total number of DWDs detected by LISA, are reported in Table \ref{tab:8yrsdetections}.
In the optimistic scenario (B1) we find a total of 215 (4684) detected CBPs (BDs), corresponding to the 0.822\% (17.913\%) of the total population of detected DWDs, and to an improvement of the 259\% (211\%) over the detections of the nominal four-year mission.
The numbers for the pessimistic scenarios, (A1) for CBPs and (A2) for BDs, are instead 8 (43) detected CBPs (BDs), corresponding to 0.02\% (0.107\%) of the total DWD population, and to an improvement of the 267\% (307\%) over the 4 year detections.

~\\
~\\
~\\
In the hypothesis of eight years of observations, LISA will be able to detect SSOs with longer period $P$ and consequently larger 
separation $a$. These are the SSO orbital parameters that present a significant improvement with respect to the four-year case, i.e. 
for which a larger range of measured values is recovered, instead of only a larger number of detections within the same 
parameter interval. We plot for comparison in Figure  \ref{fig:8yrs_distributions} the distributions (injected and recovered) 
of these two quantities for both time frames of eight and four years.
In general the longer the LISA observational period, the longer the SSO period and separation that will be recovered.
This can be easily visualised in Figure \ref{fig:8yrs_distributions} where the eight-year bulk of detected CBPs (BDs), 
presents periods up to $\sim$12 ($\sim$30) yr, compared to only $\sim$6 ($\sim$10) years over a four-year mission.
A similar trend is observed for the separation $a$, as of course this is directly related to the period.

Similarly, Figure  \ref{fig:sepVmass} shows the mass $M$ of the detected SSOs (B1 scenario), as a function of the 
semi-major axis $a$. 
Both detections obtained in a four- and eight-year time frame are shown next to each other for comparison.
We note that during eight-year observations LISA will generally be able to identify a larger number of lighter exoplanets below 2 $\Mj$.
This is because a longer baseline would allow us to disentangle the gravitational pull of the small exoplanet from the gravitational waveform, and consequently consent to measure $K$ and $P$ with a relative precision better than 30$\%$.
Similarly, the SSO range of detectable separations $a$ are roughly doubled in an eight-year mission.
Such an increase in the parameter space enables LISA to be more compatible with imaging surveys, but also with the bulk of radial velocity surveys, for which a good overlap is already visible during the nominal mission.
During both the four-year and eight-year surveys there is no real comparison with the bulk of the transit population, but this is barely a feature of the constructed SSO population, which we limited at 0.1 au. Synergies are possible between 0.1 au and 1 au, however.
 
\subsection{Error distributions of third-body parameters} 
\label{sec:errordistribution}

In this subsection we look at the distributions of 1$\sigma$ errors for the parameters $K$ and $P$, i.e.~the third-body parameters which are interesting from an observational perspective.
We report first the best and average error with which these parameters are recovered for detectable systems, i.e.~for systems which already have relative errors on both $K$ and $P$ estimated to be below 30\%.
Again we focus on the optimistic, median, and pessimistic scenarios, as selected above.
We only analyse data collected with a nominal four-year LISA mission.

In the CBP optimistic (B1), median (C1), and pessimistic (A1) scenarios, we obtain an average relative error on $K$ of the 14.9\%, 14.9\%, and 21.0\%, respectively.
The best recovered $K$ values are instead measured with relative errors of 0.86\%, 2.1\%, and 10.0\%, respectively, for the same three models above.
The average relative errors on $P$ are instead 4.4\%, 10.9\%, and 16.5\% for the three models (B1), (C1), and (A1), respectively.
The best recovered $P$ values are measured with an estimated relative error of 0.040\%, 0.23\%, and 9.6\%, again, respectively, for the same models mentioned above.

For BDs we obtain instead an average relative error on $K$ of 12.0\%, 12.3\%, and 11.8\%, for the optimistic (B1), median (B2), and pessimistic (A2) scenarios, respectively.
The same parameter is recovered with a best relative error of 0.10\%, 0.16\%, and 3.4\%, respectively, in the same three scenarios.
The average relative errors on $P$ are instead 3.3\%, 3.8\%, and 5.3\%, while the best recovered relative errors are 0.0049\%, 0.013\%, and 0.28\%, again for the the scenarios (B1), (B2), and (A2), respectively.

\section{Discussion}
\label{sec:discussion}

The results presented above show that during the four years of its nominal mission, LISA will be able to detect from a few to a few tens of CBPs down to a few Jupiter masses and up to a few astronomical units in separation.
Analogously we find that LISA will likely detect from several to few thousands BDs in roughly the same semi-major axis range.
These observations will be of fundamental importance for the field of exoplanetary science.
As shown in Figure \ref{fig:detectionsGalaxy} in the optimistic scenario LISA detections will be distributed all over the Milky Way, but even in a pessimistic scenario we would be able to detect at least some exoplanet far outside the solar neighbourhood.
In our study we only considered a Galactic population of DWDs, but we stress that LISA will be able to observe DWDs even in nearby galaxies \citep[e.g.~in the Magellanic Clouds and M31; ][]{Korol2018} and consequently, if conditions are optimal (e.g.~high values of S/N, $f_0$, $M$, ...), it could also detect  extragalactic bound CBPs/BDs \citep{TamaniniDanielski2019}, possibly leading to the discovery of the first bound extragalactic SSO.
Meanwhile, expanding the exoplanetary census beyond the local Galactic environment with GW observations, will help integrate the information collected (and that will be collected) with current (and future) EM 
surveys, and it will provide a more robust and unbiased statistic on the life of giant exoplanets. 
If this population is not detected, given the mass-separation parameter space accessible to LISA, we can confidently say that SSO do not survive a second CE phase and are either destroyed or ejected from the system. 
But whether or not the population exists, beyond the pure survival rates we will  
set constraints on the dynamical evolution of the tertiary body consequent 
to the CE phases and the binary mass ejections. 
A more robust statistic will also allow us to have a better understanding on the existence and nature of planetary generations, by testing the dynamical stability timescale of the systems and identifying if any correlation between the orbital properties of the systems is present. 
Inevitably, if the range of parameters detected is large, for instance if exoplanets are both found orbiting 
short (within the maximum radius of the CE of the progenitors), and wide orbits (where giant planets usually form), depending on the binary cooling time we could gather information on both formation and migration processes. 
The same reasoning applies to the BDs, for which these further studies would help address their difference  
from planets.

Our results also suggest that an extended LISA mission, up to eight years, will yield a larger 
parameter space than the one spanned by the nominal four-year mission, and a more robust statistic.
The number of detected CBPs and BDs will more than double, implying an incremental trend which grows more than linearly.
This is mainly because a longer observational window allows us to unlock the detections of SSOs with longer period, as clearly shown in Figure  \ref{fig:8yrs_distributions}.
To give a numerical example we note that, in the B1 scenario, over four years LISA will detect 0.32\% (8.48\%) of DWDs with a CBPs (BDs), while over eight years it will detect 0.82\% (17.91\%) of them (cf.~Tables \ref{tab:detections} and \ref{tab:8yrsdetections}).
This clearly shows that a higher percentage of the underlining SSO population will be detected with a longer observational time period, providing another scientific case for an extension of the LISA mission beyond its nominal four-year lifetime.
If the maximal envisaged duration of the mission is considered, namely ten years, then the results within our optimistic scenario suggest that LISA should be able to detect more than $\sim$250 new CBPs and more than $\sim$6000 new BDs.

For our analysis we tested detection rates of single SSOs (1 $\Mearth  < M < 0.08\ \Msun$) orbiting DWDs with the future LISA mission.
We assumed circular orbits that satisfy the stability criteria by \citet{HolmanWiegert1999} and no mass
transfer among the two WDs.
Galactic DWDs represent a stellar population older than $100\,$Myr, ergo they are not expected to follow the spiral structure of the Milky Way (see Figure \ref{fig:detectionsGalaxy}). 
For this reason, in our fiducial  simulation we neglected  the spiral structure itself and we distributed binaries in a smooth exponential disc potential with a prominent central bulge. Even when adopting a high resolution numerical simulation for the mass distribution, the contrast between the spiral arms and the background disc in GWs is too low to be detected (Wilhelm et al. in prep.).
We also note that the space density of DWDs in the solar neighbourhood is three orders of magnitude lower than that of main sequence stars \citep[e.g.][]{hol18}. This translates into a low detection statistics when comparing GW detection with currently used EM methods for the detection of exoplanets (see Figure \ref{fig:detectionsGalaxy}, top right panel). However, because the GW signal scales as $1/d$ instead of $1/d^2$, which is typical of EM observations, exoplanets can be detected farther away than will ever be possible at optical wavelengths (out to the far side of the Milky Way and satellite galaxies e.g.  the Magellanic Clouds; \citealt{Korol2018,Korol2019}).

The core composition of WDs could be a relevant element with respect to the O.R. of second generation exoplanets. Given the enrichment with heavy elements of the envelope of CO-progenitors, occurring at the end of the asymptotic giant branch, planets should be more frequent around compact CO-core DWDs (than around DWDs with a He-core WD primary;  \citealt{ZorotovicSchreiber2013}). The CO WDs should have much higher metallicities than He WDs, which is a characteristic that 
in a protoplanetary disc would promote the formation of a greater number of high-mass giant planets \citep{JohnsonLi2012}.
In our mock population 38\% of DWDs are He - He type, 27\% are He - CO, 33\% are CO - CO, and 2\% are CO - ONe (see \citealt{too12} for more details and formalism). We note however that these percentages depend on the adopted stellar evolution tracks in the binary population synthesis and can differ significantly from one model to another. 
We also note that the absolute magnitudes of DWDs in our mock population are derived from the WD cooling curves of pure hydrogen atmosphere model of \citet{Holberg2006}. Thus, by construction all binaries in the simulation are composed of DA\footnote{In the spectral classifications of WDs DA stands for hydrogen-dominated atmospheres.} WDs.

\begin{figure*}[hp!]
    \centering
    \includegraphics[width=0.9\textwidth]{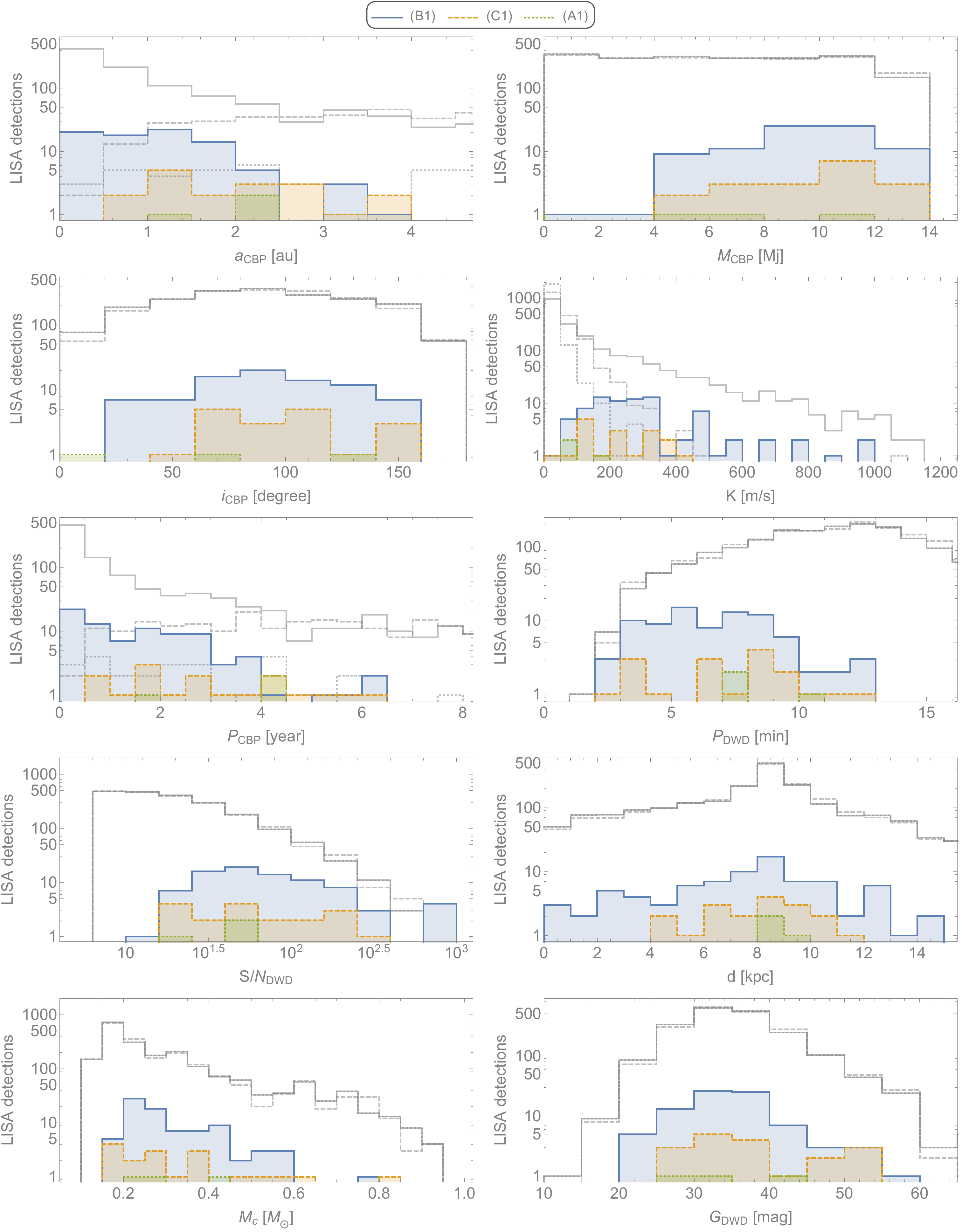}
    \captionof{figure}{
    Injected vs. detected population distributions for CBPs and its hosts in the 
    optimistic (B1, solid), intermediate (C1, dashed), and pessimistic (A1, dotted) scenarios (cf.~Table~\ref{tab:detections}).
    The injected population distribution of the three scenarios
    is shown in grey for comparison.
    From top to bottom and left to right: semi-major axis,  mass, inclination, $K$, planetary period, DWD period (denoted as $P_{\rm DWD}$ rather than $P_b$), S/N of the DWD (denoted as S/N$_{\rm DWD}$ rather than S/N), system distance, chirp mass, total $Gaia$ G magnitude of the two WDs.
    } 
    \label{fig:CBP_distributions_log}
\end{figure*}

\begin{figure*}[!p]
    \centering
    \includegraphics[width=0.9\textwidth]{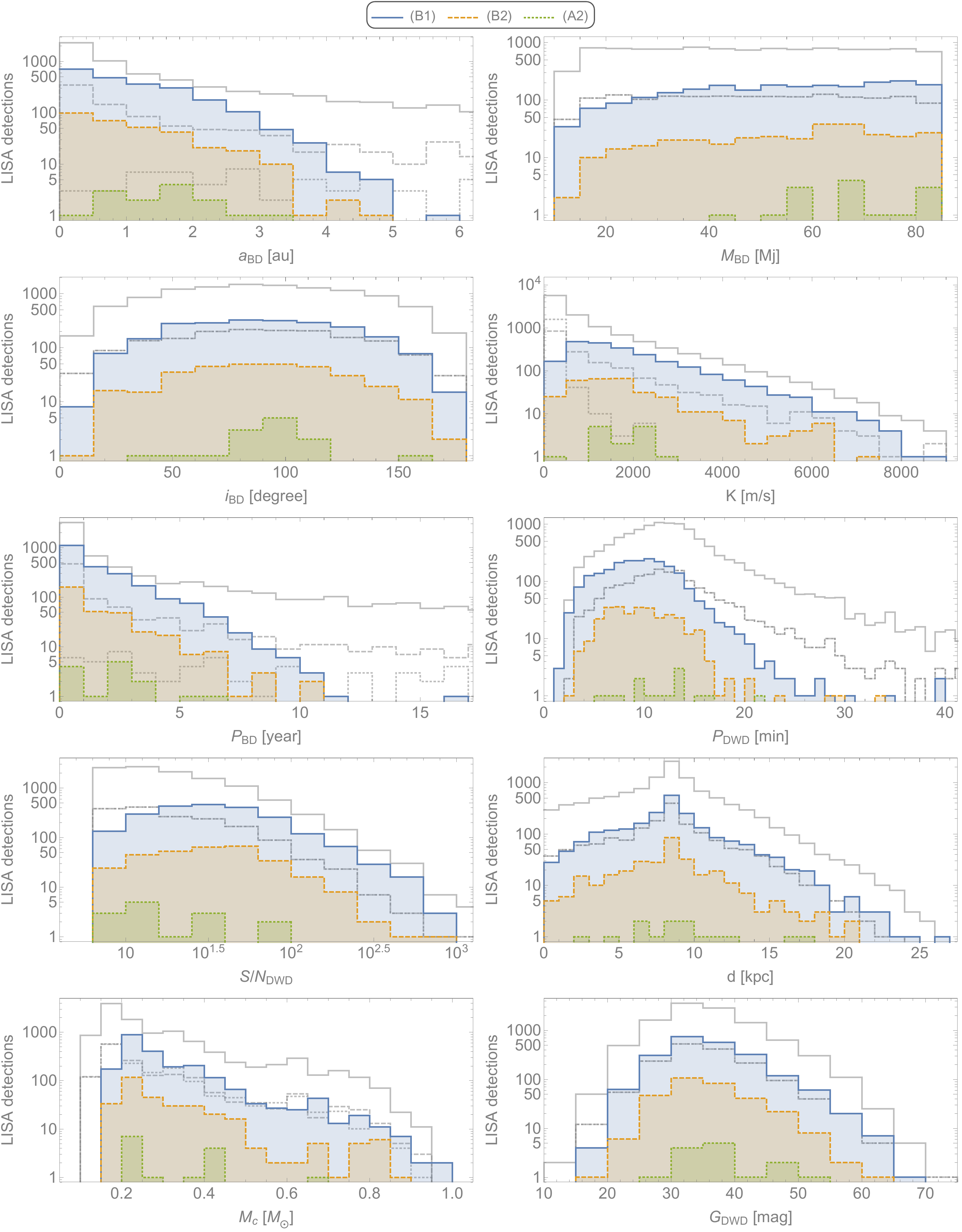}
    \caption{
    Injected vs. detected population distributions for BDs and its hosts in the 
    optimistic (B1, solid), intermediate (B2, dashed), and pessimistic (A2, dotted) scenarios (cf.~Table~\ref{tab:detections}). The injected population distribution of the three scenarios is shown in grey for comparison.
    From top to bottom and left to right: BD semi-major axis, mass, inclination, $K$, BD period, DWD period (denoted as $P_{\rm DWD}$ rather than $P_b$), S/N of the DWD (denoted as S/N$_{\rm DWD}$ rather than S/N), system distance, chirp mass, total $Gaia$ G magnitude of the two WDs.
    }
    \label{fig:BD_distributions_log}
\end{figure*}

As an example of LISA capabilities, in Table \ref{tab:specificcases} we report the parameters of the system with the least massive planet detected in this analysis, 
the system with the highest signal-to-noise of the DWDs (S/N), and the system with the longest planetary period,  for which the sky localisation error boxes measure 1.77\arcmin$^2$, 0.14\arcmin$^2$, and 12.6\arcmin$^2$, respectively.
As suggested by Table \ref{tab:specificcases} and confirmed by Figure \ref{fig:CBP_distributions_log}, binaries with $P_b$ < 10 min are optimal for detecting circumbinary companions. This result was expected because low orbital period DWDs emit high frequencies GWs; it is thus easier to discern the Doppler perturbation in the GW waveform produced by the circumbinary object \citep{TamaniniDanielski2019}.
Moreover, the higher the S/N, the easier it is to detect the same perturbations, 
meaning that detections of both CBPs and BDs are biased towards DWDs with high S/N and low orbital period (within the global DWD population that LISA will observe). 
This can be quickly confirmed by Figures \ref{fig:CBP_distributions_log} and \ref{fig:BD_distributions_log} simply by comparing the recovered versus the injected distributions of both S/N and $P_{\rm b}$. 
Furthermore high S/N necessarily corresponds to high frequency DWDs for two reasons: the GW amplitude scales as $f^{2/3}$ and LISA is more sensitive at $f \sim10^{-2}$ Hz, where we find DWDs with shortest periods (a few minutes; \citealt{LISAcallpaper}). 
Because of this, high frequency and high S/N sources can be detected anywhere in the Galaxy (as shown in Figure \ref{fig:MWinGWs}), 
with a peak at $\sim 8.5\,$kpc due to the high density of DWD in the bulge (Figure \ref{fig:BD_distributions_log}).\\
With reference to short period binaries, the time $\tau$ these bodies will take before colliding
can be approximated by
\begin{equation}
    \tau \simeq 1 \,{\rm Myr} \  \left( \frac{P_b}{12\,{\rm min}} \right)^{8/3} \left( \frac{{M_c}}{0.3\,{\rm M}_{\odot}} \right)^{-5/3}
\end{equation}
meaning that, for a $P_b \leq$ 10 min and a typical chirp mass $M_c$ = 0.2 $\Msun$ \citep{Korol2017, TamaniniDanielski2019}, 
the colliding time is $\tau \lesssim 1.558$ Myr.
Even considering the maximum chirp mass of a DWD detectable by LISA, say 1 $\Msun$, and its minimum orbital period, say 3 min, the DWD will not merge before 3300 years.

\begin{table*}[t]
\centering
\resizebox{\textwidth}{!}{%
    \begin{tabular}{c|c|c|c|c|c|c|c|c|c|c|c|c}
        \hline
         & S/N$_{\rm DWD}$ & $d$ [kpc] &
         $m_1$ [$\Msun$] & $m_2$ [$\Msun$] & $P_b$[min] & $i_b$ [deg] & 
         $M$ [$\Mj$] & $a$ [au] & $i$ [deg] & $P$ [yr] &
         $\Delta\Omega$ [arcmin$^2$]
         & $K$ [m/s]\\
         \hline
         x) & 290 & 8.12 &
         0.53585 & 0.53146 & 4.29 & 26.8 & 
         0.27 & 2.22 & 119.33 & 3.19 &
         1.77
         & 4.3\\
         y) & 963 & 1.55 & 
         0.74955 & 0.47068 &  5.25 & 119.8 & 
         10.99 & 0.61 & 130.79 & 0.43 &
         0.14
         &  272.30\\
         z) & 182 & 6.35 &
         0.32285 & 0.30066 & 2.66 & 42.22 & 
         11.11 & 3.67 & 138.36 & 8.82 &
         12.6
         & 137.11 \\
         \hline
    \end{tabular}%
    }
    \caption{Among the detected SSOs in all 4 yr scenarios we report the systems with the least massive CBP detected (x, A2), with the highest S/N (y, B1), and with the CBP with longest period (z, C1).
    In this table the DWD S/N is denoted as S/N$_{\rm DWD}$ rather than S/N as in the text.    }
    \label{tab:specificcases}
\end{table*}

Our detections present some events far in the tails of the observed distributions.
These events are usually associated with a combination of high DWD S/N, high DWD GW frequency, and high SSO mass, which correspond to a stronger perturbation in the GW signal and which are thus easier to detect.
Consequently it is not surprising that they can be detected even for unusual values of the SSO parameters.
The CPB period distribution in Figure \ref{fig:CBP_distributions_log} for example shows few events with periods around six years,
while the bulk of the distribution is set on periods shorter than four years. 
A numerical example in the C1 case is given by the system with the longest detected planetary period (8.82 yr)
which report a S/N = 182, 
a GW frequency $f_0 = $ 12.53 mHz, $M = 11.11\ \Mj$, and $a = $ 3.67 au. 
This is even more evident for BDs, for which in the D1 scenario we detect a system with a BD whose orbital period is 19.5 yr,  even if the global detections are set at $P$ < 12 yr (see Figure \ref{fig:BD_distributions_log}, where an outlier with $P\sim 17$ yr is also appearing in the (B1) scenario). This event is characterised by S/N = 169
$f_0 =$ 12.37 mHz, $M =$ 78.86 $\Mj$, and $a =$7.2 au, which shows that such outliers can only be detected for systems with high S/N, high frequency, and high SSO mass.
All this shows that, with ideal conditions, LISA could detect CBPs with periods up to $P \sim 10$ yr and BDs with periods up to $P \sim$ 20 yr, with only four-year observations.  
The bulk of detections however are expected at $P < 4$ yr for CBPs and at $P < 11$ yr for BDs (see Figure  \ref{fig:CBP_distributions_log} and \ref{fig:BD_distributions_log}); only rare events appear at higher orbital periods.
Moreover, as noted by \citet{TamaniniDanielski2019}, the Fisher matrix approach adopted in this work might not be reliable for events with extremely high S/N and further more detailed data analysis techniques should be used to determine the real detectability of such systems.

In this work we accounted for only one circumbinary SSO for DWD, but observations show that multiple SSOs can orbit evolved binaries, i.e. NN Ser (b,c), UZ For (b,c), or  HU Aqr, which is a system that hosts one giant planet and two BDs \citep{Gozwski2015}. 
Consequently, since multiple circumbinary objects could co-exist (see \citealt{VerasGansicke2015} for a single WD case), our results report lower limits of detections in all the possible scenarios of mass and planet-to-binary separation distributions.
We note however that additional SSOs, or even a low-mass star, orbiting the same DWD would complicate the GW signal detected by LISA because of the simultaneous Doppler perturbations of different circumbinary objects. 
This might worsen the precision with which the SSO orbital parameters are recovered, possibly leading to some detections being missed.
Future analyses, which lie outside the scope of the present work, will be needed to explore the detectability of CBPs and BDs in systems with multiple orbiting SSOs or with a third star composing hierarchical triples with the DWD. 
Alongside with it, also studies on the dynamical stability of multi-planets, similarly to e.g.  \citealt{Mustill2014, VerasGansicke2015, Kostov2016, Mustill2018}, but specific for SSO orbiting DWDs are needed to understand the dynamical grounds of these objects. 
This might need to take into account also the possibility of co-existing generations of planets, aspect that 
would necessarily make the analysis computationally expensive.

We mentioned in Sec. \ref{sec:injection} that no specific O.R. for planets orbiting a binary WD is available, therefore
we used the atmospheric pollution frequency (25-50\%) for single stars, robustly measured by \cite{Zuckerman2010} and \cite{Koester2014}. 
Recently \cite{Wilson2019}, using Spitzer and Hubble data, estimated the pollution rate in
WDs in wide binaries to be 67$^{+10}_{-15}$ $\%$, consistent within 2$\sigma$ to the single WDs 
value measured in the same work  (45$\pm$4$\%$) and to the rate value applied in our study.
These rates are also consistent to the O.R. of planets transiting single WDs measured by \cite{VanSluijs2018}, using K2 data. For Jupiter-size planets and planetary periods between 10.12 - 40 days, the authors calculated a detection probability of 53.3$\pm$3.0$\%$ within a 68$\%$ confidence interval. 
Given that the length of the survey was only 40 days, there is no information for larger periods and extrapolation at wider orbits would be highly inaccurate given the lack of physical constraints. 
We consequently decided the 50$\%$ upper hand limit reported by \citet{Koester2014} to be our reference value.
The total number of detections in a scenario with 25$\%$ O.R. (lower hand limit) 
can be directly estimated from our results in Table \ref{tab:detections} since all numbers scale linearly and can thus just be divided by 2. Specifically, in the optimistic scenario (B1) CBPs and BDs detections would be 42 (0.16\%) and 
1109 (4.24\%), respectively.
In the CBP pessimistic scenario (A1) instead we would only get one detection in four years of observations, while in the pessimistic BD scenario (A2) we would count seven detections.
The same reasoning applies to the eight-year results, for which numbers in Table \ref{tab:8yrsdetections} should just be halved. 

Concerning BDs detections, we notice that in the optimistic scenario (B1) they amount to $\sim$27 times the number of CBPs (versus the 2.3 times factor for the pessimistic A2 case), representing the 8.48$\%$ of the 
binaries DWD population in the Milky Way. Such a  result was expected given that, assuming the same mass of the binary, a more massive object would produce a larger motion of centre of mass of the three-body system (cf.~Eq.~\eqref{eq:K}), and hence a larger shift in frequency, which is easier to detect.
The mass distributions in Figure \ref{fig:CBP_distributions_log} and \ref{fig:BD_distributions_log} show this dependency very clearly.
The residuals of the injected versus detected population of BDs, normalised to the injected population, in the B1 case, i.e.~that presenting a more robust statistics, goes from 91\% for BD masses between 15-20 $\Mj$ to 73\% for BD masses between 75-80 $\Mj$.
On average thus SSOs with larger masses have a higher probability of being detected by LISA, as expected.
The total BDs (i.e. over the mass range $M$ > 13 $\Mj$) normalised residuals (80\%) are indeed smaller that the total CBPs ($M \leq$ 13 $\Mj$)  normalised residuals (96\%),  again for the optimistic (B1) scenario.\\
Besides, GWs do not allow for a direct measurement of the mass of the SSO. The mass can be estimated only once 
both $K$ and $P$ are known, and only after we assume a value (or a range of values) for both the binary mass ratio, 
and the SSO orbital inclination $i$, in analogy with radial velocity measurements  \citep{TamaniniDanielski2019}.
These considerations imply that without EM counterpart data it will be difficult to discern a CBP from a BD for masses around 13 $\Mj$, especially if the GW measurement is not precise enough;  the needed level of precision depends on the specific case.
Only an independent EM estimation of the binary total mass, the SSO orbital inclination, and the SSO radius will enable us to unambiguously characterise the nature of the GW detected SSO (see \ref{sec:future}). 

In this investigation we injected SSOs with masses up to 0.08 $\Msun$.
This was justified by the fact that the separation between the nature of planets and BDs is still uncertain. 
By applying the same reasoning the WD pollution, whose O.R. we used, could be also driven by low-mass BDs, i.e. very massive exoplanets. 
Because of this we took into account the largest possible mass range to cover both populations. 
However, had we assumed that the O.R.~was only valid for planetary masses ($M \leq$ 13 $\Mj$), i.e. by abruptly excluding the hypothesis that pollution could be also caused by objects able to at least burn deuterium, the CBPs detection rates would have been higher. 
On the other hand we would have not detected BDs, as none of them would have been injected.

\subsection{The LISA duty cycle and system identification}

Our study was based on the nominal LISA mission lifetime, i.e. four years of uninterrupted observations.
However, during $\sim$ 30\% of this time, LISA will not be acquiring scientific data because of expected maintenance operations (duty cycle).
Nevertheless, even though the total effective observational period will be below four years, a periodic stop of scientific operation should not negatively impact the detection capability of LISA, at least for long-living GW sources if additional data analysis tools are employed \citep{Baghi:2019eqo}.
Our results should thus not be affected by the duty cycle of LISA, albeit we note that a future dedicated investigation is required to address this aspect fully.

We expect DWDs to be very numerous in the Milky Way. Population synthesis studies predict that $\sim 10^6$ DWDs have periods within the LISA frequency band, \citep[e.g.][]{Korol2019}. Only 1\% of these objects will have S/N > 7 and be individually resolved by LISA, while the overlapping signals of the remaining DWDs will sum up to form a Galactic noise background. 
Using the same mock DWD population as in this work (Figure \ref{fig:MWinGWs}), \citet{lit19} show that the DWD confusion background 
is mostly confined between $\sim 0.4$ and $\sim 4\,$mHz, meaning that at frequencies $ f >4\,$mHz DWDs 
can be individually identified.
The typical LISA error on sky localisation for DWDs is $<10\,$deg$^2$ \citep{Korol2019}, although for DWDs with a detected SSOs, which we recall are biased towards higher frequency and higher S/N, the error is much lower.
For example in our optimistic scenario (B1) the mean sky location accuracy of the DWDs with a detected CBP (BD) is 0.29 (2.83) deg$^2$. The 74.5$\%$ (37.4$\%$) of these systems are above $f \sim$ 4 mHz.
This implies that LISA DWDs with a detected CBP/BD have higher chances to be spotted by EM telescopes.

\subsection{What does the future look like?}
\label{sec:future}

Mainly because they are intrinsically faint and physically small, DWDs are difficult targets for optical
telescopes. Typically, spectra of DWDs are virtually identical to those of single WDs, while their eclipses are very short \citep[e.g.][]{reb19}.
This drastically limits the observed volume, with the most distant detached DWD detected around $\sim 2.4\,$kpc \citep[][]{bur19}. 
Nonetheless, the number of DWDs detected with EM techniques are expected to increase substantially with the upcoming future all-sky and wide optical surveys, which also cover low Galactic latitudes, for example~BlackGem \citep{bloem15}, GOTO \citep[Gravitational-wave  Optical  Transient Observer;][]{ste17}, Gaia \citep{gaia16}, and LSST. 
There can be different detection strategies to identify the EM counterparts. 
For example, high cadence (several observations per night) photometric surveys such as ZTF \citep[Zwicky Transient Facility;][]{bel14} can be used to search for variable sources with an orbital period provided by LISA.
Surveys with longer cadence (one observation in a few days) such as PTF \cite[Palomar Transient Factory;][]{law09} or LSST, can also be used for finding DWD EM counterparts. 
However, in the latter case it will be important to account for the orbital period derivative (also provided by LISA if the system is chirping) caused by GW radiation; see example of retrieving J153932.16+502738.8 in PTF archival data in \cite{bur19}. 
The work by \cite{Korol2017} in particular showed that at least 100 DWDs are expected to have GW counterparts.
These predictions give us optimistic prospects for observing exoplanets and BDs around DWDs (but also other stellar compact object binaries), which would require monitoring DWDs for a few years.

Concerning a detection of a SSO signal with EM techniques we refer to the Methods section of \citet{TamaniniDanielski2019} for specific discussion on the EM synergies with GWs. 
We stress though that upcoming data from $Gaia$ will highly increase the sample of giant planets in long-period orbits around binaries with FGK-dwarf primaries, located within 200 pc from the Sun \citep{Sahlmann2015}. $Gaia$ will also detect tens or hundreds of planets ($M$ > $\sim$ 1 $\Mj$) around single WD  that, combined with those that the PLATO 2.0 mission is also expected to find \citep{PLATO}, will increase this population statistic (should they exist in the favourable region of parameter space).
This will allow us to begin placing limits on the masses of planets that can survive stellar evolution. 
Furthermore, $Gaia$, LSST and WFIRST will help detect free-floaters i.e. planets not bound to any star(s), which may help constrain the fraction of ejected planets due to mass loss \citep{Veras2014, Veras2016}. 
These observations, combined with a continuous development of long-term dynamical evolution models of planetary systems, 
will help to acquire a more focused picture on the surviving life of exoplanets.
Similarly, new studies on formation of second/third generation bodies orbiting post-CE binaries, as well as accretion studies 
of first and second generation objects, are important to address both the rates and orbital characteristics of the population 
investigated in this work and to understand whether the presence of a `surviving generation(s)' inhibits or promotes the formation process of new planets.

\subsection{Possibly planet detection by LISA around ZTFJ1539+5027}

The recently discovered ZTF J153932.16+502738.8, which has an orbital period of $\sim7\,$min at a distance $d$ = 2.34 kpc,  is a great example of the potential of the multi-messenger observations with the aforementioned surveys together with LISA \citep{bur19}.
According to our detection definition in Sec \ref{sec:method} we note that a planet with mass $M$ = 1 $\Mj$ orbiting J153932.16+502738.8 at a separation of 1 au, would not be detected by LISA.
Using the measured orbital parameters of J153932.16+502738.8 \citep{bur19,lit19}, setting $\Psi_0 = 0$ (initial DWD orbital phase), marginalising over $\psi_b$ (the polarisation angle unconstrained by EM observations), 
and assuming the planetary orbit to have $\varphi_0=\pi/2$ and $i = \pi/2$ (most favourable orientation), 
LISA would measure its parameters as $K = 31.4 \pm 39.4$ m/s (relative accuracy: 126\%), $P = 1.1043 \pm 0.0857$ yr (7.8\%) and $\varphi_0 = \pi/2 \pm 1.08$ rad (69\%), where the sky location has been fixed to the real one measured for J153932.16+502738.8. 
We see that, although the planetary period is well constrained, $K$ (and $\varphi_0$) are unconstrained, 
we would thus not be able to estimate the mass of the planet.
We note however that an accurate measure of the planetary period could be extremely useful if taken in combination with other EM observations, which for J153932.16+502738.8 can be easily planned.
This highlights the multi-messenger potential of LISA in terms of exoplanetary observations.

The situation changes for more massive planets.
If we consider a planet with 13 $\Mj$ at the same separation of 1 au and repeat the analysis above, 
we find that LISA would be able to measure $K = 405.1 \pm 38.4$ m/s (9.5\%), 
$P = 1.0967 \pm 0.0067$ yr (0.61\%), and $\varphi_0 = \pi/2 \pm 0.0853$ rad (5.4\%).
In this case the planet would be easily detected by LISA with accurate measurements of its orbital parameters.
This also implies that any BD orbiting at the same separation from J153932.16+502738.8 would be detected by LISA and its parameters would be measured with even higher precision.

If instead J153932.16+502738.8 would appear at a distance of 9 kpc (implying S/N$\simeq$37), well beyond the Galactic bulk where the majority of DWDs are expected to be detected by LISA, we would only be able to measure the same 13 $\Mj$ planet with a relative accuracy of $\sigma_K / K = 96\%$ and $\sigma_P / P = 10\%$, which again shows that, even if we will not access any information on the mass of the planet, we would still be able to measure its orbital period quite well. 
However in this case EM complementary observations will be impossible to obtain with the current instrumentation.

We finally compute the probability that an SSO in our optimistic (B1) population has the same $f_0$, $f_1$, and $\eta$ of J153932.16+502738.8 (practically within 10\% of these values).
Among all 13086 SSOs present in our population, only 14 (0.11\%) have these characteristics, 12 of which are BDs and 2 CBPs.
Of the 12 BDs, 5 are detectable by LISA while the 2 CBPs are not detectable.
If we project these numbers on J153932.16+502738.8, and we recall that we are assuming a 50\% O.R., we find that this system has a 17.9\% probability of harbouring BDs detectable by LISA.
The same reasoning cannot be performed for CBPs for which we can only conclude that the probability that J153932.16+502738.8 has a circumbinary planet detectable by LISA is very small.

\section{Summary}
\label{sec:conclusions}

In this work we quantitatively estimated the detection rates, by the\ LISA\ mission led by ESA, of circumbinary SSOs (i.e. planets and BDs) orbiting Galactic detached DWDs. 
To do so we injected a simulated population of SSOs into a synthetic population of already formed DWDs, with an O.R. of 50$\%$ i.e. the observed frequency of polluted WDs \citep{Koester2014}.
We then applied the method presented by \citet{TamaniniDanielski2019} to probe how many systems we can identify to have a SSO perturbing the DWD GW signal because of its gravitational pull. 
Given that currently no theoretical and observational constraints are present to define such a specific population,
we tested various combination of semi-major axis and mass distributions for estimating the number of detections over the course of the LISA nominal mission.
Our analysis identified an optimistic and pessimistic scenario for which we counted a total of 63 (2218), and 3 (14) detections of circumbinary exoplanets (BDs) in the Milky Way, respectively.
These numbers corresponds to 0.317$\%$ (8.482$\%$), and 0.011$\%$ (0.054$\%$) of the total DWDs visible by LISA, and these have more than doubled in a time frame of eight-year continuous LISA observations, corresponding to a realistic extended mission.
In such a case the range of recovered planetary periods (and semi-major axis) would double for planets and increase almost threefold for BDs.
The SSO detections that we found are also biased towards high frequency and high S/N binaries, as expected from basic considerations.
The advantages of using the GW method for detection of CBPs and BDs comes from the fact that GWs are not affected by dust extinction and can be measured from all over the Milky Way and the Local Group.
In constrast to EM techniques, this method is most efficient in the most dense regions of the Milky Way like the central bulge.
A full investigation of a realistic observational strategy, including EM complementary observations, 
will be performed in future studies.

\begin{acknowledgements}
      We would like to thank Pierre-Olivier Lagage, Vincent Van Eylen, and Dimitri Veras, for useful discussions.
      C.D.~acknowledges support from the LabEx P2IO, the French ANR contract 05-BLAN-NT09-573739.
      V.K.~and E.M.R.~thank the NWO WARP Program, grant NWO 648.003004 APP-GW.
      V.K.~acknowledges support from the Netherlands Research Council NWO (Rubicon Science grant 019.183EN.015).
      This work was supported by the COST action CA16104 “Gravitational waves, black holes and fundamental physics” (GWverse) through a Short Term Scientific Mission (STSM) awarded to N.T. COST actions are supported by COST (European Cooperation in Science and Technology).
      This work made use of the python package $mw\_plot$ (\url{https://pypi.org/project/mw-plot/}) and of the annotated image of the Milky Way by NASA/JPL-Caltech/R.Hurt (SSC/Caltech).
\end{acknowledgements}

\bibliographystyle{aa} 
\bibliography{Biblio}

\end{document}